\documentclass[showpacs,preprintnumbers,amsmath,amssymb,sublabel,pra,twocolumn]{revtex4}
\usepackage{amsfonts}
\usepackage{amssymb}
\usepackage{amsmath}
\usepackage{amsthm}
\usepackage{mathrsfs}
\usepackage{bm}
\usepackage{graphicx}
\usepackage{dcolumn}
\usepackage{multirow}
\usepackage{float}
\usepackage{sublabel}
\usepackage{times,epsfig}

\begin{document}
\title{General quantum key distribution in higher dimension}

\author{Zhao-Xi Xiong$^{1,3}$, Han-Duo Shi$^1$,
Yi-Nan Wang$^1$, Li Jing$^1$, Jin Lei$^1$,
Liang-Zhu Mu$^1$\footnote{muliangzhu@pku.edu.cn}, and Heng
Fan$^2$\footnote{hfan@iphy.ac.cn}}
\affiliation{
$^1$School of Physics, Peking University, Beijing 100871, China\\
$^2$Institute of Physics, Chinese Academy of Sciences, Beijing
100190, China\\
$^3$Department of Physics, Massachusetts Institute of Technology, Cambridge,
Massachusetts 02139, USA
}
\date{\today}

\begin{abstract}
We study a general quantum key distribution protocol in higher dimension. In this protocol, quantum states in arbitrary $g+1$ ($1\le g\le d$) out of all $d+1$ mutually unbiased bases in a d-dimensional system can be used for the key encoding. This provides a natural generalization of the quantum key distribution in higher dimension and recovers the previously known results for $g=1$ and $d$. In our investigation, we study Eve's attack by two slightly different approaches. One is considering the optimal cloner for Eve, and the other, defined as the optimal attack, is maximizing Eve's information. We derive results for both approaches and show the deviation of the optimal cloner from the optimal attack. With our systematic investigation of the quantum key distribution protocols in higher dimension, one may balance the security gain and the implementation cost by changing the number of bases in the key encoding. As a side product, we also prove the equivalency between the optimal phase covariant quantum cloning machine and the optimal cloner for the $g=d-1$ quantum key distribution.
\end{abstract}


\maketitle

\section{INTRODUCTION}

Quantum key distribution (QKD) is a promising application of quantum mechanics. The first QKD protocol was proposed by Bennett and Brassard in 1984 (BB84) \cite{BB84} and has been proved to be unconditionally secure \cite{LoChau,Shor}. This protocol was later generalized to the six-state protocol \cite{Bruss}.
Meanwhile, various other protocols were developed, among which, for example, was the Ekert 91 protocol \cite{ekert}. In the past decades, significant progress has been made both theoretically and experimentally in establishing point-to-point as well as network types of key distributions; see, for example, Refs. \cite{review,pangroup,grangier,shields,pengcz,yamomoto}.

In a simple two dimensional system, one can either use four quantum states or six quantum states, which correspond to the BB84 protocol and the six-state protocol respectively, to encode binary symbols. In a higher-dimensional system of dimension $d$ \cite{prime}, there are altogether $d+1$ mutually unbiased bases (MUBs) available for the QKD. The counterparts of the BB84 protocol and the six-state protocol of a $d$-dimensional system are the $2$-basis protocol and the $(d+1)$-basis protocol. Naturally, one may think of a general $(g+1)$-basis protocol ($g=1,~2,~...~,~d$), where arbitrary $g+1$ MUBs are utilized to encode d-ary symbols. In the past few years, there have been many studies on higher-dimensional QKD protocols. The QKD using four-level systems were done first \cite{4dim2000,4dim2003}, and interests soon extended to the $d$-dimensional case \cite{QKDquditBell,CerfSecurity,classicalAdvtg,entangQuditScrty,2basisCohUlt,2basis2wayErrorTol,2wayGen2007,symext,g1dCohScrtyPrf}. Most of these studies, however, focused on the $2$-basis and the $(d+1)$-basis cases, and the research on the most general case is still absent. In this article, we present the study on such a general $(g+1)$-basis QKD protocol.

In principle, the quantum states of a higher dimension can be encoded in a continuous variable system such as a harmonic oscillator \cite{gottesman}. It is of fundamental interest, theoretically, how a $(g+1)$-basis QKD protocol is formalized. Our investigation of the $(g+1)$-basis protocol is a natural generalization of the previously studied cases. The results of our systematic study may help one balance the security gain and the implementation cost by changing the choice of $g$.

In this article, our general QKD protocol that uses arbitrary $g+1$ MUBs is the following. Suppose Alice, the sender, wants to send Bob, the receiver, a set of classical symbols consisting of $0$, $1$, ..., $d-1$. To start, Alice encodes every symbol, say $i$, into a pure quantum state $|i\rangle$ or $|\tilde i^{(k)}\rangle$ of one randomly chosen MUB out of the $g+1$ and sends the state to Bob. Upon receipt, Bob measures the state using again a randomly chosen basis, which is correct with probability $1/(g+1)$. Subsequently, the choice of bases are publicly announced by Alice, and the states measured in bases different from they are prepared are discarded by the two parties. In the absence of any eavesdropping and environmental noises, Alice and Bob are then left with identical strings of symbols while they are left with partially correlated strings in the presence of Eve, an eavesdropper. By checking the agreement of a subset of the symbol sequence, Alice and Bob can decide whether to continue or abort the protocol. If the disagreement is below a threshold, they then perform a direct reconciliation and a privacy amplification to obtain a set of shared key. In this article, we consider that Eve attacks the QKD by intercepting and cloning the state being sent to Bob. For simplicity, we think of that Eve uses a fixed and balanced (balanced between different bases) cloning transformation for each qudit, and that Eve measures her state before the one-way post-processing between Alice and Bob.

In this article, we investigate Eve's attack scheme by two slightly different approaches, both starting from a general form of cloning transformation proposed in Ref. \cite{CerfAsymmetric}. One approach is considering the optimal cloner for Eve where we maximize the fidelities of the state of Eve. The other is considering maximizing the information Eve has about Alice's state, which, rather than maximizing the fidelities, is defined as the optimal attack. We do these separately in Sec. \ref{seccloner} and Sec. \ref{secattack} and compare the two approaches subsequently. As we shall see, they give different results. The second approach is done in a somewhat restrictive sense, but it is sufficient to prove the difference between the optimal cloner and the optimal attack. Sec. \ref{secanalytical} gives some analytical solutions to the optimal cloner in special cases, including the symmetric cloner corresponding to a general $g$. In Sec. \ref{secphase}, we introduce a side product of our first approach to the QKD attack, where we present the link between a QKD cloner and a revised asymmetric form of the optimal symmetric phase covariant quantum cloning machine proposed in Ref. \cite{FanPhase} and prove the optimality of the latter. In Sec. \ref{secconclusion}, we end the article by a brief conclusion.

\section{THE OPTIMAL CLONER OF EVE\label{seccloner}}

Now we investigate the optimal cloner that Eve can use. Before proceeding, let us first introduce some notations. In dimension $d$, there are $d+1$ mutually unbiased bases, namely $\{|i\rangle\}$ and $\{|\tilde{i}^{(k)}\rangle\}$ ($k=0,~1~,~...~,~d-1$), which more explicitly are
\begin{eqnarray}
|\tilde{i}^{(k)}\rangle=\frac{1}{\sqrt{d}}\sum_{j=0}^{d-1}\omega^{i(d-j)-ks_j}|j\rangle,
\end{eqnarray}
with $s_j=j+...+(d-1)$ and $\omega=e^{i\frac{2\pi}{d}}$ \cite{ANewProof}. By saying two bases, say $\{|\tilde{i}^{(0)}\rangle\}$ and $\{|\tilde{i}^{(1)}\rangle\}$, are mutually unbiased, we mean that $|\langle \tilde{i}^{(1)}|\tilde{k}^{(0)}\rangle |=\frac {1}{\sqrt {d}}$ for any $|\tilde{k}^{(0)}\rangle$ and $|\tilde{i}^{(1)}\rangle$ in the two bases respectively. The generalized Pauli matrices $\sigma_x$ and $\sigma_z$ act on the states so that $\sigma_x|j\rangle=|j+1\rangle$ and
$\sigma_z|j\rangle=\omega^j|j\rangle$. Throughout the article, we omit the ``modulo $d$," which is the case here. We define $U_{mn}=\sigma_x^m\sigma_z^n$ so that $U_{mn}|j\rangle=\omega^{jn}|j+m\rangle$. Finally, the generalized $d$-dimensional Bell states read
\begin{eqnarray}
|\Phi_{mn}\rangle=(\mathbb{I}\otimes U_{m,-n})|\Phi_{00}\rangle,
\end{eqnarray}
with $m,~n=0,~1,~...~,~d-1$ and
\begin{equation}
|\Phi_{00}\rangle=\frac{1}{\sqrt{d}}\sum_{j=0}^{d-1}|j\rangle|j\rangle.
\end{equation}

Now, we consider the optimal cloner for Eve. Suppose a state $|\psi\rangle_A$ is sent by Alice and intercepted by Eve. Then, Eve prepares a maximally entangled state $|\Phi_{00}\rangle_{E'E}$ and performs a unitary transformation $U$ of the general form proposed in Ref. \cite{CerfAsymmetric}:
\begin{equation}
U=\sum_{m,n=0}^{d-1}a_{mn}(U_{mn}\otimes U_{m,-n}\otimes \mathbb{I}).\label{cloningtransform}
\end{equation}
Here, $a_{mn}$ are the parameters of the unitary transformation, satisfying $\sum_{m,n}|a_{mn}|^2=1$. This transformation yields
\sublabon{equation}
\begin{eqnarray}
&&U|\psi\rangle_A|\Phi_{00}\rangle_{E'E}\nonumber\\
&&=\sum_{m,n}a_{mn}U_{mn}|\psi\rangle_B\otimes|\Phi_{-m,n}\rangle_{E'E} \label{Psiouta} \\
&&=\sum_{m,n}b_{mn}|\Phi_{-m,n}\rangle_{BE'}\otimes U_{mn}|\psi\rangle_E, \label{Psioutb}
\end{eqnarray}
\sublaboff{equation}
where $A$, $B$, $E$ and $E'$ denote Alice, Bob, Eve and her cloning machine respectively. $b_{mn}$ are the discrete fourier transform of $a_{mn}$, i.e. $b_{mn}=\frac{1}{d}\sum_{k,r}a_{kr}\omega^{kn-rm}$. Eqs. (\ref{Psiouta}) and (\ref{Psioutb}) make it convenient to write down the density matrices of Bob as well as Eve. For $|\psi\rangle_A$ being state $|i\rangle$ or $|\tilde i^{(k)}\rangle$ of each MUB, we have
\sublabon{equation}
\begin{eqnarray}
\rho_B&=&\sum_{m,n=0}^{d-1}|a_{mn}|^2|i+m\rangle\langle i+m|,\label{rhoB}\\
\tilde{\rho}^{(k)}_B&=&\sum_{m,n=0}^{d-1}|a_{mn}|^2(U_{mn}|\tilde{i}^{(k)}\rangle_B)(\langle\tilde{i}^{(k)}|_B U_{mn}^\dagger),\label{rhoBk}\\
\rho_E&=&\sum_{m,n=0}^{d-1}|b_{mn}|^2|i+m\rangle\langle i+m|,\label{rhoE}\\
\tilde{\rho}^{(k)}_E&=&\sum_{m,n=0}^{d-1}|b_{mn}|^2(U_{mn}|\tilde{i}^{(k)}\rangle_E)(\langle\tilde{i}^{(k)}|_E U_{mn}^\dagger)\label{rhoEk}\\
&&~~~~~~~~~~~~~~~~~~~~~~~~~~~~~~(k=0,~1,~...~,~g-1).\nonumber
\end{eqnarray}
\sublaboff{equation}

Let us consider the fidelities of the states $B$ and $E$ with respect to all the $g+1$ different bases. The fidelity here can be defined as $F\equiv \langle\psi|\rho_{red.}^{out}|\psi\rangle$, the value of which differs for states of different bases. With the help of the properties of the Pauli matrices and the Bell states, we figure out the fidelities of $B$ and $E$ for each mutually unbiased basis:
\sublabon{equation}
\begin{eqnarray}
F_B&=&\sum_n|a_{0n}|^2,\label{FB}\\
\tilde{F}^{(k)}_B&=&\sum_m|a_{m,km}|^2,\label{FBk}\\
F_E&=&\frac{1}{d}\sum_m|\sum_n a_{mn}|^2,\label{FE}\\
\tilde{F}_E^{(k)}&=&\frac{1}{d}\sum_n|\sum_m a_{m,n+km}|^2\label{FEk}\\
&&(k=0,~1,~...~,~g-1).\nonumber
\end{eqnarray}
\sublaboff{equation}

For the QKD using $g+1$ MUBs, without loss of generality, we suppose that the bases $\{|i\rangle\}$, $\{|\tilde i^{(0)}\rangle\}$, ..., $\{|\tilde i^{(g-1)}\rangle\}$ are chosen by the two legitimate parties. For simplicity, we assume that Eve's attack is balanced, i.e. she induces an equal probability of error for all the $g+1$ MUBs. This assumption follows from the reasoning that Eve can be detected easily by unbalanced disturbance otherwise, and that, as we find, Eve cannot maximize all her $g+1$ fidelities simultaneously if the disturbance is unbalanced. Hence, we assume
\begin{equation}
F_B=\tilde{F}^{(0)}_B=...=\tilde{F}^{(g-1)}_B.\label{balance}
\end{equation}
At this point, to obtain an optimal cloner, Eve has to maximize her fidelities for a given $F_B$, which quantifies the
disturbance. We claim and will show later that Eve can maximize all her $g+1$ fidelities simultaneously and that they are equal. We start by a ``vectorization" of the matrix elements of $(a_{mn})$. Let
\begin{eqnarray}
\vec \alpha_i&=&(a_{1,1i},...,a_{d-1,(d-1)i})~~(i=0,~...~,~g-1),\label{vecalphaidef}\\
\vec A&=&(A_1,~...~,A_{d-1}),\label{vecAdef}\\
A_i&=&\sum_{j\not=0,i,...,(g-1)i}^{d-1}a_{ij}~~~~(i=1,~...~,d-1).\label{vecAidef}
\end{eqnarray}
The rest elements are $a_{0j}$ ($j=0,~...~,d-1$). Eq. (\ref{balance}) gives the following restrictions:
\begin{eqnarray}
&&\sum_{j=1}^{d-1}|a_{0j}|^2=F_B-|a_{00}|^2,\label{restriction1}\\
&&||\vec\alpha_i||^2=F_B-|a_{00}|^2~~~~(i=0,...,g-1).\label{restriction2}
\end{eqnarray}
One of Eve's fidelity $F_E$ now reads
\begin{eqnarray}
F_E=\frac{1}{d}(|\sum_{j=0}^{d-1}a_{0j}|^2+||\sum_{i=0}^{g-1}\vec\alpha_i+\vec A||^2).\label{FEvec}
\end{eqnarray}
Eqs. (\ref{restriction1})-(\ref{FEvec}) tell us how the fidelities of Eve and Bob are coupled with each other.

Now we show how Eqs. (\ref{vecalphaidef})-(\ref{FEvec}) work by doing the $g=2$ version of them. The results for a generic $g$ can be obtained analogously. For $g=2$,
\begin{eqnarray*}
&&F_E=\frac{1}{d}(|\sum_{j=0}^{d-1}a_{0j}|^2+||\vec\alpha_0+\vec\alpha_1+\vec A||^2),\\
&&\sum_{j=1}^{d-1}|a_{0j}|^2=||\vec\alpha_0||^2=||\vec\alpha_1||^2=F_B-|a_{00}|^2.
\end{eqnarray*}
We tentatively fix the values of $|a_{00}|$ and $\vec A$. By vector manipulation, it is easy to find that the maximum $F_E$ is achieved when
\begin{eqnarray*}
&&\vec\alpha_0,~\vec\alpha_1\propto\vec A,\\
&&a_{01}=...=a_{0,d-1}=\sqrt{\frac{F_B-|a_{00}|^2}{d-1}}e^{iArg(a_{00})}.
\end{eqnarray*}
Unfixing $|a_{00}|$ and $\vec A$, $F_E$ becomes a function of them and is positively correlated to $||\vec A||$ directly. Using the normalization condition $\sum_{m,n}|a_{mn}|^2=1$, we find that $||\vec A||$ is maximal when $a_{i,2i}=a_{i,3i}=...=a_{i,(d-1)i}$ ($i=1,~...~,~d-1$). Now we visualize the matrix $(a_{mn})$:
\begin{eqnarray*}
(a_{mn})=
\left(
\begin{array}{ccccc}
v&x&x&x&...\\
x_1&x_1&y_1&y_1&\ldots\\
x_2&y_2&x_2&y_2&\ldots\\
x_3&y_3&y_3&x_3&\ldots\\
\vdots&\vdots&\vdots&\vdots&\ddots
\end{array}
\right),
\end{eqnarray*}
where
\begin{eqnarray*}
x&=&\sqrt{\frac{F_B-|v|^2}{d-1}}e^{iArg(v)},\\
x_i&=&\sqrt{F_B-|v|^2}\frac{A_i}{||\vec A||},~~~y_i=\frac{A_i}{d-2},\\
||\vec A||&=&\sqrt{1+2|v|^2-3F_B}.
\end{eqnarray*}
Correspondingly, Eve's fidelity $F_E$ becomes
\begin{multline*}
F_E=\frac{1}{d}[(|v|+\sqrt{(d-1)(F_B-|v|^2)})^2\\
+(\sqrt{1+2|v|^2-3F_B}+2\sqrt{F_B-|v|^2})^2].
\end{multline*}
Thus far, we have maximized one of Eve's fidelities. The other fidelities can be obtained simply by transposing the roles of ``the horizontal direction," ``the vertical direction," and ``the diagonal direction" of the matrix $(a_{mn})$ and redefining $\vec\alpha_0$, $\vec\alpha_1$, and $\vec A$ accordingly. Doing so, we find that the condition for optimization of all Eve's fidelities, with an overall phase omitted, is
\begin{eqnarray*}
(a_{mn})=
\left(
\begin{array}{ccccc}
v&x&x&x&\ldots\\
x&x&y&y&\ldots\\
x&y&x&y&\ldots\\
x&y&y&x&\ldots\\
\vdots&\vdots&\vdots&\vdots&\ddots
\end{array}
\right),\label{supa_mn2}
\end{eqnarray*}
where the elements are all real, and
\begin{eqnarray*}
x=\sqrt{\frac{F_B-v^2}{d-1}},~~~~y=\sqrt{\frac{1+2v^2-3F_B}{(d-1)(d-2)}}.
\end{eqnarray*}
All Eve's fidelities are found to be equal. They are equal to
\begin{eqnarray*}
F_E=\frac{1}{d}\{[v+(d-1)x]^2+(d-1)[2x+(d-2)y]^2\}.
\end{eqnarray*}

The results for a generic $g$ are analogous to those for $g=2$. The difference is that some ``2" are substituted by ``g." We thus present the following optimization condition and the optimal Eve's fidelities:
\begin{eqnarray}
&&F_E=\tilde{F}_E^{(0)}=...=\tilde{F}_E^{(g-1)}=\nonumber\\
&&\frac{1}{d}\{[v+(d-1)x]^2+(d-1)[gx+(d-g)y]^2\},\label{gfidelity}\\
&&a_{mn}=
\begin{cases}
v,~m=n=0,\\
x,~m=0,n\neq0~or~m\neq0,n=km,\\
y,~otherwise
\end{cases}\label{gmatrix}\\
&&(k=0,~...~,~g-1),~~~~~~~~~~~~~~~~~~~~~~~for~some~v,\nonumber
\end{eqnarray}
where
\begin{eqnarray}
x=\sqrt{\frac{F_B-v^2}{d-1}},~~~~y=\sqrt{\frac{1+gv^2-(g+1)F_B}{(d-1)(d-g)}}.\label{gxy}
\end{eqnarray}

One notices that there is still an undetermined, independent variable $v$. To achieve the maximal $F_E$ for a given $F_B$, one needs to further optimize the value of $v$. In some cases, this can be done analytically, but, in the general case, there seems to be no evident analytical expression. Those analytical results are presented in Sec. III. We do numerical calculation for the general case.

\begin{figure}[!h]
\includegraphics[width=8cm]{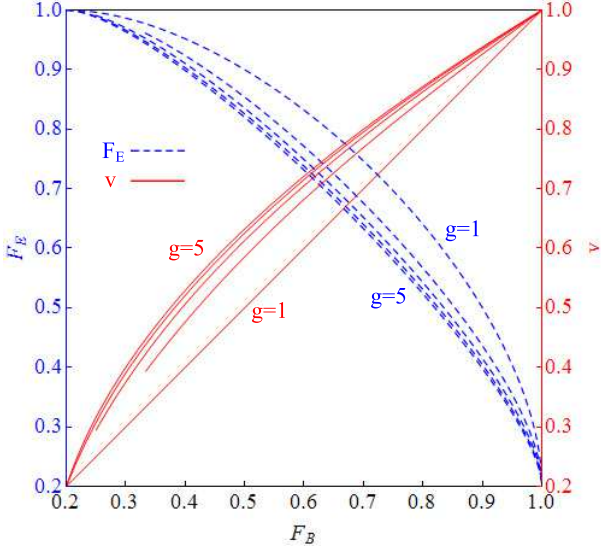}
\caption{(color online). The curves of the fidelity $F_E$ (dashed blue lines) and the parameter $v$ (solid red lines) as functions of $F_B$ of the optimal cloner for $d=5$ and $g=1,2,...,5$. The $g=1$ and $g=5$ lines are derived from Eqs. (\ref{1fidelity}) and (\ref{dfidelity}) respectively. The curves for general $g$'s are numerically computed from Eqs. (\ref{gfidelity})-(\ref{gxy}). The general-$g$ curves in some domains with very small $F_B$ are incomputable by Eqs. (\ref{gfidelity})-(\ref{gxy}), but they can be obtained by exchanging the roles of $E$ and $B$.}
\label{FIG1}
\end{figure}

To show the performance of the optimal cloner, we choose $d=5$ as an example and plot the optimized $F_E$ and $v$ curves as functions of $F_B$ in FIG. \ref{FIG1}.
The figure shows that, as $F_B$ increases, $F_E$ decreases, i.e. as the disturbance decreases, the state $E$ resembles the original state less. The figure also shows reasonable shifts of the curves as $g$ varies. As $g$ increases, i.e. as more bases are used, the $F_E$ curve shifts down, suggesting that $F_E$ is lower for a given $F_B$. The difference between adjacent curves is small for large $g$'s.

\section{ANALYTICAL SOLUTIONS TO THE OPTIMAL CLONER\label{secanalytical}}

In this section, we present the analytical expressions of $F_E$ and $v$ for some special cases. The first one is the $g=1$ case. For $g=1$, an easier method gives that
\begin{eqnarray}
v=F_B,~~x=\sqrt{\frac{F_B(1-F_B)}{d-1}},~~y=\frac{1-F_B}{d-1},\label{1vxy}
\end{eqnarray}
whence
\begin{eqnarray}
F_E=\frac1d[\sqrt{F_B}+\sqrt{(d-1)(1-F_B)}]^2.\label{1fidelity}
\end{eqnarray}
The second set of analytical results is for $g=d$. In this case, there exists no $y$ in the matrix $(a_{mn})$,  and $v$ has a fixed value given below.
\begin{eqnarray}
v=\sqrt{\frac{(d+1)F_B-1}{d}},~~x=\sqrt{\frac{1-F_B}{d(d-1)}}\label{dvxy}
\end{eqnarray}
whence
\begin{eqnarray}
F_E&=&\frac{1}{d^2}\left[\sqrt{(d+1)F_B-1}+\sqrt{(d-1)(1-F_B)}\right]^2\nonumber\\
&+&(1-F_B).\label{dfidelity}
\end{eqnarray}
Eqs. (\ref{1fidelity}) and (\ref{dfidelity}) recover the previous known results in Ref. \cite{CerfSecurity}. Here, they are related by parameters $v$ and $g$ within a unified framework.

The third case in which we have analytical results is that Eve and Bob have equal fidelities, i.e. $F_E=F_B=F$. We start by observing the following fact: When $F_B=F_E=F$, Eq. (\ref{gfidelity}) is equivalent to
\begin{multline}
\sqrt{(d-g)[1+gv^2-(g+1)F]}\\
=v\sqrt{d-1}-(g+1)\sqrt{F-v^2}.
\end{multline}
We square this equation and let $u=\sqrt{F-v^2}$, and we end up with an quadratic equation of $u$ and $v$,
\begin{multline}
\frac{(d+1)(g+1)}{d-g}u^2+\frac{2d-g-1}{d-g}v^2\\
-\frac{2(g+1)\sqrt{d-1}}{d-g}uv=1,\label{2ordereqn}
\end{multline}
which represents an ellipse centering at the origin. The fidelity $F=u^2+v^2$ is the square of the distance from the origin to the point $(u,v)$ and is thus maximal at one end of the major axis. It is easy to find that point by diagonalizing the coefficient matrix. The eigenvalues of that coefficient matrix are found to be
\begin{eqnarray}
\lambda_\pm=\frac{d}{2(d-g)}[(g+3)\pm P(d,g)],\label{symeigenv}
\end{eqnarray}
where
\begin{equation}
P(d,g)=\sqrt{(g+3)^2-8\frac{(d-g)(g+1)}{d}}.
\end{equation}
By further calculating the eigenvectors, we figure out the maximal fidelity $F$ and the corresponding parameter $v$. The maximal fidelity is
\begin{eqnarray}
F&=&\frac{1}{\lambda_-}=\frac{2}{d}\frac{d-g}{(g+3)-P(d,g)}.\nonumber\\
&=&\frac{2}{d}\frac{d-g}{(g+3)-\sqrt{(g+3)^2-8\frac{(d-g)(g+1)}{d}}}\label{gsymmfidelity}.
\end{eqnarray}
The corresponding $v$ satisfies
\begin{eqnarray}
\frac{u}{v}=\frac{(d+1)(g+1)-(d-g)\lambda_+}{-(g+1)\sqrt{d-1}}
\end{eqnarray}
and is thus
\begin{multline}
v=[\frac{2}{d}\times\frac{(d-1)(d-g)}{(g+3)-P(d,g)}\times\\
\frac{1}{(d-1)+[\frac{d}{2(g+1)}P(d,g)-1-\frac{d}{2}\frac{g-1}{g+1}]^2}]^\frac12.\label{gsymv}
\end{multline}
We remark that Eqs. (\ref{gsymmfidelity}) and (\ref{gsymv}) are the results for an optimal symmetric cloning machine that clones arbitrary $g+1$ MUBs and that has not been studied. We can find that, as $g$ increase, $F$ increases as it is expected.

\section{THE OPTIMAL ATTACK OF EVE\label{secattack}}

In our $g+1$ basis QKD protocol, we consider that Eve intercepts each state and copies it using a fixed cloning machine of the form of Eq. (\ref{cloningtransform}). We now think of that Eve's scheme is to maximizes her information about the state, rather than the fidelity, for a given, balanced disturbance. We consider that the post-processing between Alice and Bob is one-way, consisting of a direct reconciliation and a privacy amplification. Therefore, the amount of secret information extractable by Alice and Bob reads
\begin{eqnarray}
r=I_{AB}-I_{AE},\label{r}
\end{eqnarray}
where $I_{AB}$ (or $I_{AE}$) is the mutual information between the two classical strings of symbols of Alice and Bob (or of Alice and Eve). Note that Eve can measure $E'$ and $E$ jointly, so $I_{AE}$ represents the mutual information between the classical random variable $A$ and the random variable pair $E'$ and $E$, which is Eve's joint-measurement outcome (for convenience, we denote the associated random variables again by $A$, $B$, $E'$, and $E$).

Let us now see what restrictions are imposed on Eve. We supposed that Eve's attack is balanced between different bases, i.e. the fidelities of Bob's state are same for different MUBs. Here, for similar reasons, we also suppose that, for each basis, the probabilities that Bob make different errors are equal (one can check that this restriction is compatible with the results in Sec. {\ref{seccloner}}). Say the error Bob makes is $m$ ($m=0,~1,~...~,~d-1$), i.e. Bob's symbol is greater than Alice's symbol by $m$. Then, from Eqs. (\ref{rhoB})-(\ref{rhoEk}), we find the following explicit expressions for these restrictions:
\begin{eqnarray}
\sum_{j=0}^{d-1}|a_{mj}|^2&=&
\begin{cases}
F_B,~~~~~~m=0,\\
\frac{1-F_B}{d-1},~~m\neq0,
\end{cases}\label{attackrest1}\\
\sum_{i=0}^{d-1}|a_{i,ki-m}|^2&=&
\begin{cases}
F_B,~~~~~~m=0,\\
\frac{1-F_B}{d-1},~~m\neq0
\end{cases}\label{attackrest2}\\
&&(k=0,~1,~...~,~g-1).\nonumber
\end{eqnarray}
Under these restrictions, one can easily find that the mutual information between Bob and Alice is given by
\begin{eqnarray}
I_{AB}=\log_2d+F_B\log_2F_B+(1-F_B)\log_2\frac{1-F_B}{d-1}\label{IAB}.
\end{eqnarray}
To find $I_{AE}$ for one basis $\{|i\rangle\}$, we first rewrite Eq. (\ref{Psiouta}) as
\begin{multline}
|i\rangle_A \rightarrow \sum_{m,j} \left(\frac{1}{\sqrt d}\sum_na_{mn}\omega^{n(i-j)}\right) |i+m\rangle_B|j\rangle_{E'}|j+m\rangle_E.\label{transformrewrite}
\end{multline}
As mentioned above, $I_{AE}$ is between the random variable $A$ and the random variable pair $(E',E)$. Suppose $(E',E)$ takes the value $(e',e)$. Eq. (\ref{transformrewrite}) tells us that it is equivalent to represent $(e',e)$ by $(m,e')$. Thus, Eve's information $I_{AE}$ can be written as
\begin{eqnarray}
I_{AE}=&-&\sum_{m,e'}p(m,e')\log_2 p(m,e')\nonumber\\
&+&\sum_{a,m,e'}p(a)p(m,e'|a)\log_2 p(m,e'|a).\label{IAE}
\end{eqnarray}
$a$ is the value the random variable $A$ takes. We assume that Alice sends symbols randomly. Thus, in Eq. (\ref{IAE}), $p(a)=\frac1d$. The other terms in Eq. (\ref{IAE}) are given below, as they are derived from Eq. (\ref{transformrewrite}).
\begin{eqnarray}
p(m,e'|a)&=&\frac1d|\sum_na_{mn}\omega^{n(a-e')}|^2,\label{pme'a}\\
p(m,e')&=&\frac1d\sum_a p(m,e'|a).\label{pme'}
\end{eqnarray}

The expressions for $I_{AE}$ of the other bases can be written analogously. Most generically, the optimal attack can be found by maximizing the $I_{AE}$'s under the restrictions of Eqs. (\ref{attackrest1}) and (\ref{attackrest2}). This can be done in principle, but it is hard because of the great number of variables and summations.
Hence, instead, we here maximize Eve's information conditionally: We suppose that the matrix ($a_{mn}$) takes the form of Eq. (\ref{gmatrix}) but that $v$ is adjustable to maximize Eve's information (rather than Eve's fidelities). One can check that Eq. (\ref{gmatrix}) satisfies Eqs. (\ref{attackrest1}) and (\ref{attackrest2}), and one need only calculate $I_{AE}$ with respect to one basis because Eq. (\ref{gmatrix}) is balanced between different bases. We shall compare the results with those of the optimal cloner approach later. In this more restrictive case, $a_{mn}$ are given partial freedom, but as we shall see, this is sufficient to prove the deviation of the optimal attack from the optimal cloner.

We refer to Eq. (\ref{gmatrix}) and find that Eq. (\ref{pme'a}) now reads
\begin{multline}
p(m,e'|a)=\\
\begin{cases}
\frac{[v+(d-1)x]^2}{d},~~~~~~~~~~m=0,~e'=a,\\
\frac{(v-x)^2}{d},~~~~~~~~~~~~~~~~~~m=0,~e'\neq a,\\
\frac{[gx+(d-g)y]^2}{d},~~~~~~~~~m\neq0,~e'=a,\\
\frac{(x-y)^2}{d}|\frac{1-\omega^{mgt}}{1-\omega^{mt}}|^2,~~m\neq0,~e'=a-t~~(t\neq0).
\end{cases}\label{pme'aCond}
\end{multline}
Then, numerical calculation can be easily done by adjusting $v$ to maximize $I_{AE}$, and both $I_{AB}$ and $I_{AE}$ become functions of $F_B$. Let us focus on the critical point where the amount of extractable information is zero, i.e., according to Eq. (\ref{r}), $I_{AB}=I_{AE}$. As usual, we substitute $F_B$ with $D_I$, the disturbance, defined as $D_I=1-F_B$. We compute the $D_I$'s associated with zero extractable information for several $d$'s and list them in TABLE \ref{TABLE1}. The values of these critical $D_I$'s show regular behaviors: As $d$ or $g$ increases, $D_I$ increases, i.e. as more bases are used, higher disturbance is acceptable for Alice and Bob.
\begin{table}[!h]
\centering
\begin{tabular}{cc|ccccccc}
\hline
\hline
\multicolumn{2}{c|}{\multirow{2}{*}{$D_{I}(\%)$}}   &\multicolumn{7}{c}{$g$}\\
                                                    \cline{3-9}
                                            &       &~1~&~2~&~3~&~4~&~5~&~6~&~7~\\
\hline
\multicolumn{1}{c|}{\multirow{4}{*}{$d$}}   &~~2~~  &~14.64~&~15.64~&&&&&\\
\multicolumn{1}{c|}{}                       &~~3~~  &~21.13~&~22.47~&~22.67~&&&&\\
\multicolumn{1}{c|}{}                       &~~5~~  &~27.60~&~28.91~&~29.12~&~29.20~&~29.23~&&\\
\multicolumn{1}{c|}{}                       &~~7~~  &~30.90~&~32.10~&~32.26~&~32.32~&~32.36~&~32.38~&~32.39~\\
\hline
\hline
\end{tabular}
\caption{
The disturbance $D_I$ associated with zero extractable information for Alice and Bob. These values are obtained with conditionally maximized $I_{AE}$, where $(a_{mn})$ is restricted to the form of Eq. (\ref{gmatrix}). For the $d$'s and $g$'s we consider, $D_I$ shows regular behaviors: Both when $d$ increases and when $g$ increases, $D_I$ increases. Since the unconditionally maximized $I_{AE}$ can be slightly higher, the real critical $D_I$ can be slightly lower than the values here.
}
\label{TABLE1}
\end{table}

It is interesting to see whether the maximizing information approach and the maximizing fidelity approach are equivalent. For definiteness, let us consider whether the optimal cloner corresponds to the maximal $I_{AE}$. We substitute the $a_{mn}$ in Eqs. (\ref{IAE})-(\ref{pme'}) by the values associated with the optimal cloner. This means that we plug into Eq. (\ref{pme'aCond}) the value of $v$ of the optimal cloner, as is calculated in Sec. \ref{seccloner}. Then, we similarly end up with a table (TABLE \ref{TABLE2}) of the disturbance associated with zero extractable information. We use a different notation $D_F$ here to indicate that it corresponds to the maximized $F_E$ rather than the maximized $I_{AE}$.
\begin{table}[!h]
\centering
\begin{tabular}{cc|ccccccc}
\hline
\hline
\multicolumn{2}{c|}{\multirow{2}{*}{$D_{F}(\%)$}}   &\multicolumn{7}{c}{$g$}\\
                                                    \cline{3-9}
                                            &       &~1~&~2~&~3~&~4~&~5~&~6~&~7~\\
\hline
\multicolumn{1}{c|}{\multirow{4}{*}{$d$}}   &~~2~~  &~14.64~&~15.64~&&&&&\\
\multicolumn{1}{c|}{}                       &~~3~~  &~21.13~&~22.99~&~22.67~&&&&\\
\multicolumn{1}{c|}{}                       &~~5~~  &~27.64~&~29.75~&~29.83~&~29.63~&~29.23~&&\\
\multicolumn{1}{c|}{}                       &~~7~~  &~31.10~&~33.24~&~33.16~&~33.00~&~32.83~&~32.64~&~32.39~\\
\hline
\hline
\end{tabular}
\caption{
The disturbance $D_F$ associated with zero extractable information, obtained simply by plugging the values of $a_{mn}$ of the optimal cloner into Eqs. (\ref{IAE})-(\ref{pme'}). $D_F$ has an irregular behavior: As $g$ increases, $D_F$ does not change monotonously. $D_F$ deviates above $D_I$ of TABLE \ref{TABLE1} except for $g=d$, in which case $v$ is fixed, and the higher the dimension, the larger the deviation. This suggests that the optimal cloner is not the optimal attack (see the text).
}
\label{TABLE2}
\end{table}
In TABLE \ref{TABLE2}, $D_F$ shows an irregular behavior: As $g$ increases, i.e. as more bases are used, $D_F$ does not always increase. $D_F$ is greater than $D_I$ except for $g=d$, in which case $v$ is fixed, and the deviation tends to be larger as $d$ increases. As we know, associated with $D_I$ is the $I_{AE}$ that is maximized under the condition that $(a_{mn})$ is of the form Eq. (\ref{gmatrix}), so the maximal $I_{AE}$ free of this condition may be slightly larger and thus the condition-free $D_I$ may be lower. Since $D_F$ is larger than the conditional $D_I$, it is larger than the condition-free $D_I$. Therefore, maximizing $F_E$ is not equivalent to maximizing $I_{AE}$, and the optimal cloner does not correspond to the optimal attack.

\section{PHASE COVARIANT QUANTUM CLONING MACHINE\label{secphase}}

We now introduce a side product of the optimal cloner approach to our $g+1$ protocol QKD. As mentioned in Ref. \cite{CerfSecurity}, the optimal cloner for $d+1$ MUBs ($g=d$) is the universal quantum cloning machine \cite{CerfAsymmetric,UQCMSecurity}. For $d$ MUBs ($g=d-1$), one may intuitively think of phase-covariant quantum cloning machine. In this section, we show that the optimal cloner of $d$ MUBs is equivalent to the optimal asymmetric phase-covariant quantum cloning machine. More specifically, we show that it is equivalent to a revised asymmetric form of the symmetric phase-covariant quantum qudit cloning machine presented in Ref. \cite{FanPhase}, and we prove the optimality of that revised form.

In Ref. \cite{FanPhase}, the following equatorial states are considered:
\begin{eqnarray}
|\Phi\rangle^{(in)}=\frac{1}{\sqrt d}\sum_{j=0}^{d-1}e^{i\phi_j}|j\rangle,\label{phasePhiin}
\end{eqnarray}
where $\phi_j$ are arbitrary phase parameters. (Thus, the corresponding MUB cloning machine should be the one that clones the bases $\{|\tilde i^{(0)}\rangle\}$, ..., and $\{|\tilde i^{(d-1)}\rangle\}$.) The explicit expression for the symmetric cloning transformation is given as
\begin{eqnarray}
|i\rangle\rightarrow \alpha|ii\rangle_{12}|i\rangle_R+\frac{\beta}{\sqrt{2(d-1)}}\sum_{j\not=i}(|ij\rangle+|ji\rangle)|j\rangle,
\label{phasesymtransform}
\end{eqnarray}
where $1$, $2$ represent the two clones while $R$ is the ancillary state. $\alpha$ and $\beta$ are real parameters that
satisfy $\alpha^2+\beta^2=1$. The optimal fidelity for the symmetric cloning machine reads
\begin{eqnarray}
F_{optimal}=\frac{1}{4d}(d+2+\sqrt{d^2+4d-4}).\label{phasesymfidelity}
\end{eqnarray}
One finds that Eq. (\ref{phasesymfidelity}) is consistent with Eq. (\ref{gsymmfidelity}).

Now we claim that the optimal asymmetric phase-covariant quantum cloning machine is equivalent to the optimal cloner of the $d$ MUBs and takes the form
\begin{eqnarray}
|i\rangle \rightarrow \alpha|ii\rangle|i\rangle+
\frac{\beta}{\sqrt{d-1}}\sum_{j\neq i}(\cos\theta|ij\rangle+\sin\theta|ji\rangle)|j\rangle,\label{phaseasymtransformation}
\end{eqnarray}
where $\theta$ is a real parameter. To prove this claim, we first write down the fidelities associated with this cloning transformation,
\begin{eqnarray}
F_1&=&\frac1d+\frac{2\alpha\beta}{d}\sqrt{d-1}\cos\theta+\frac{\beta^2(d-2)}{d}\cos^2\theta,
\label{phaseasymfidelity1}\\
F_2&=&\frac1d+\frac{2\alpha\beta}{d}\sqrt{d-1}\sin\theta+\frac{\beta^2(d-2)}{d}\sin^2\theta,
\label{phaseasymfidelity2}
\end{eqnarray}
where the constraint $\alpha^2+\beta^2=1$ still holds. We perform a numerical calculation that manipulates $\alpha$, $\beta$, as well as $\theta$ to maximize one fidelity given the other. The results show that the optimized fidelities for this asymmetric phase-covariant cloning quantum machine are equal to the fidelities of the optimal $d$-MUB cloner, as are computed in Sec. \ref{seccloner}. Since cloning equatorial states has a higher requirement than cloning $d$ MUBs, the optimality of a $d$-MUB cloner infers the optimality of an asymmetric phase-covariant quantum cloning machine with the same achieved fidelities, and their equivalency. This proves our claim.

\section{CONCLUSION\label{secconclusion}}

In this article, we study the general, d-dimensional QKD that uses arbitrary $g+1$ MUBs, focusing on the individual attack by Eve and the one-way post-processing (a direct reconciliation plus a privacy amplification) by Alice and Bob. This investigation of the general $g+1$ MUB QKD protocol is a natural generalization of the QKD in higher dimension and may help one balance the gain and the cost of the implementation. In this article, we investigate Eve's attack by two different approaches. One is maximizing $F_E$, the fidelity of Eve's state $E$, while the other is maximizing $I_{AE}$, the information Eve has about Alice classical symbol. In the first approach (Sec. \ref{seccloner}), we derive the fidelities and the parameter of the optimal cloner and demonstrate their behaviors, which are reasonable. It turns out that in some special cases, the most significant of which is the symmetric cloning, the optimal cloner can by solved analytically (Sec. \ref{secanalytical}). In the second approach (Sec. \ref{secattack}), we give the equations for the most generic calculation for $I_{AE}$ maximization, but, considering its complexity, we do instead a more restrictive version. Though restricted, the calculation still shows interesting results. In particular, it proves (except for $g=d$) the deviation of the optimal cloner from the optimal attack. Sec. \ref{secphase} is dedicated to a side product of our optimal cloner approach. We show that the optimal asymmetric phase covariant quantum cloning machine is equivalent to the optimal cloner of $d$ MUBs ($g=d-1$). We also show that this optimal asymmetric phase covariant quantum cloning machine can be formulated as a revised version of the optimal symmetric cloning transformation presented in Ref. \cite{FanPhase}. As the bottom line, we here remark that there still exist several possible extensions, which may be of future interests, for our general, $(g+1)$-basis, qudit-based QKD protocol: extension to two-way post-processing, to prime-power dimensional systems, and to the coherent attack.

\section*{ACKNOWLEDGEMENT}

We thank Xin-Quan Chen for useful discussions. This work is supported by NSFC (10974247), ``973'' program (2010CB922904) and NFFTBS (J1030310).


\begin{thebibliography}{99}
\bibitem{prime}We only consider the case that $d$ is prime.
\end{thebibliography}
\end{document}